\documentclass{article}

\usepackage[english]{babel}

\usepackage[letterpaper,top=2cm,bottom=2cm,left=3cm,right=3cm,marginparwidth=1.75cm]{geometry}

\usepackage{amsmath}
\usepackage{graphicx}
\usepackage[colorlinks=true, allcolors=blue]{hyperref}

\title{Smallest vertical beam sizes achieved in high-energy accelerators}
\author{Rogelio Tom\'as Garc\'ia$^1$, Yukiyoshi Ohnishi$^2$, and Frank Zimmermann$^1$}
\date{
$^1$ CERN, Esplanade de particules 1, 1211 Geneva 23, Switzerland.\\
$^2$ High Energy Accelerator Research Organization (KEK), Oho 1-1, Tsukuba 305-0801, Japan.\\$ $ \\
\today
}
\begin{document}
\maketitle

\begin{abstract}
FCC--ee aims at colliding electrons and positrons with vertical rms beam sizes between 30 and 50~nm with beam energies extending from about 40 GeV to 182.5~GeV. In this report, we collect experimental results of different high energy accelerator 
projects with comparable design vertical beam sizes.
\end{abstract}

\section{Smallest vertical beam sizes over time}
The Final Focus Test Beam (FFTB) in SLAC~\cite{FFTBdesign} and the Accelerator Test Facility 2 in KEK~\cite{ATF2design} were conceived to demonstrate different aspects of the Final Focus Systems (FFS) of TeV-scale e$^+$e$^-$ linear colliders as ILC~\cite{ILC} or CLIC~\cite{CLIC}. 
The  measurement of the nanometric vertical beam size in FFTB and ATF2 is performed by scanning the e$^-$ beam across a laser interference fringe pattern~\cite{shintake}. 

SuperKEKB in KEK~\cite{SuperKEKBdesign} is a circular e$^+$e$^-$ collider delivering luminosity to the Belle II experiment.

\begin{figure}[h]
\centering
\includegraphics[width=1\linewidth]{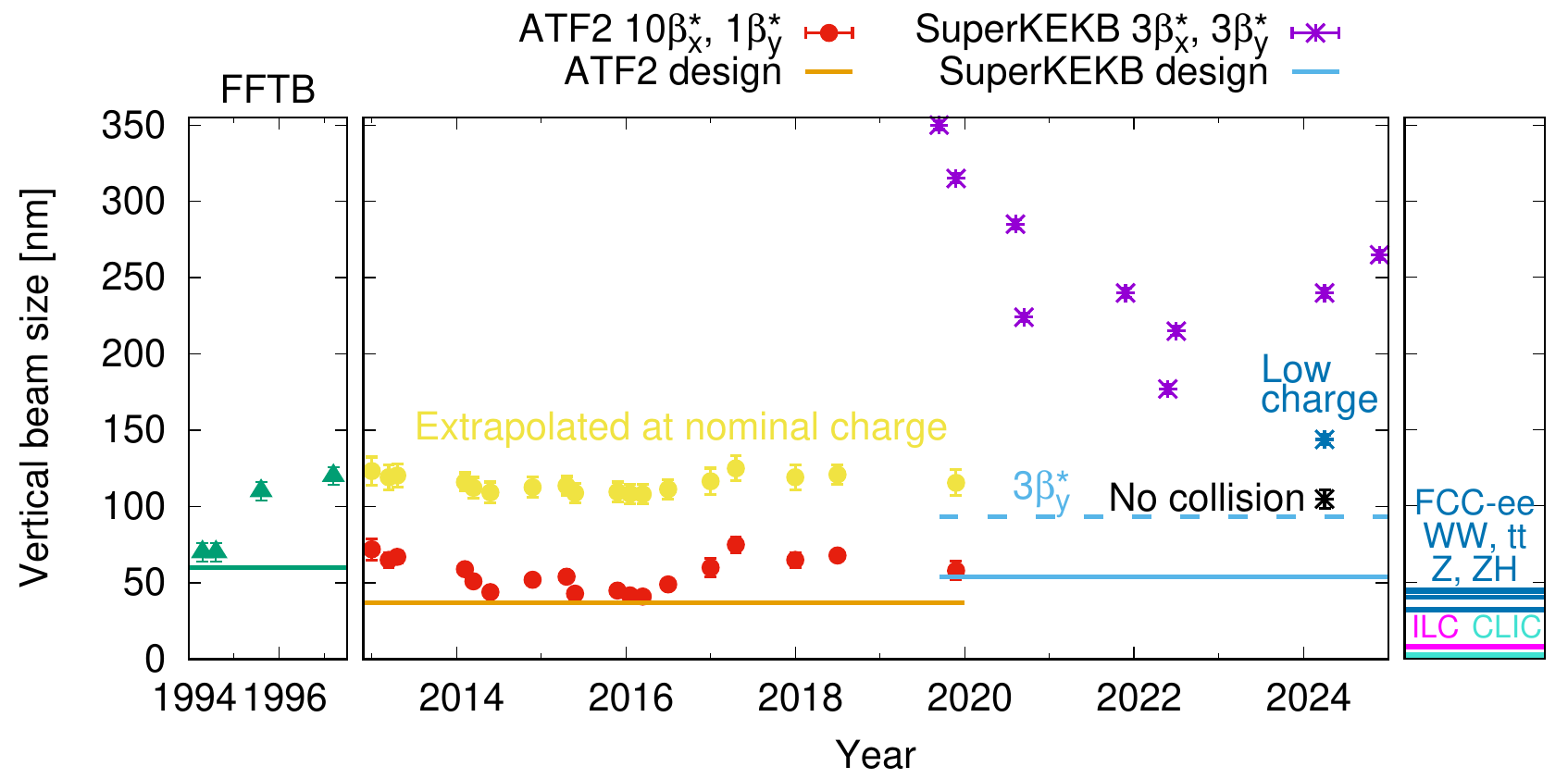}
\caption{\label{fig:sigmay} Design, measured or extrapolated vertical beam size versus time for FFTB (green), ATF2 (orange, red and yellow), SuperKEKB (blue, violet and black), FCC--ee (dark blue), ILC (magenta) and CLIC (turquoise). }
\end{figure}

\begin{figure}[h]
\centering
\includegraphics[width=1\linewidth]{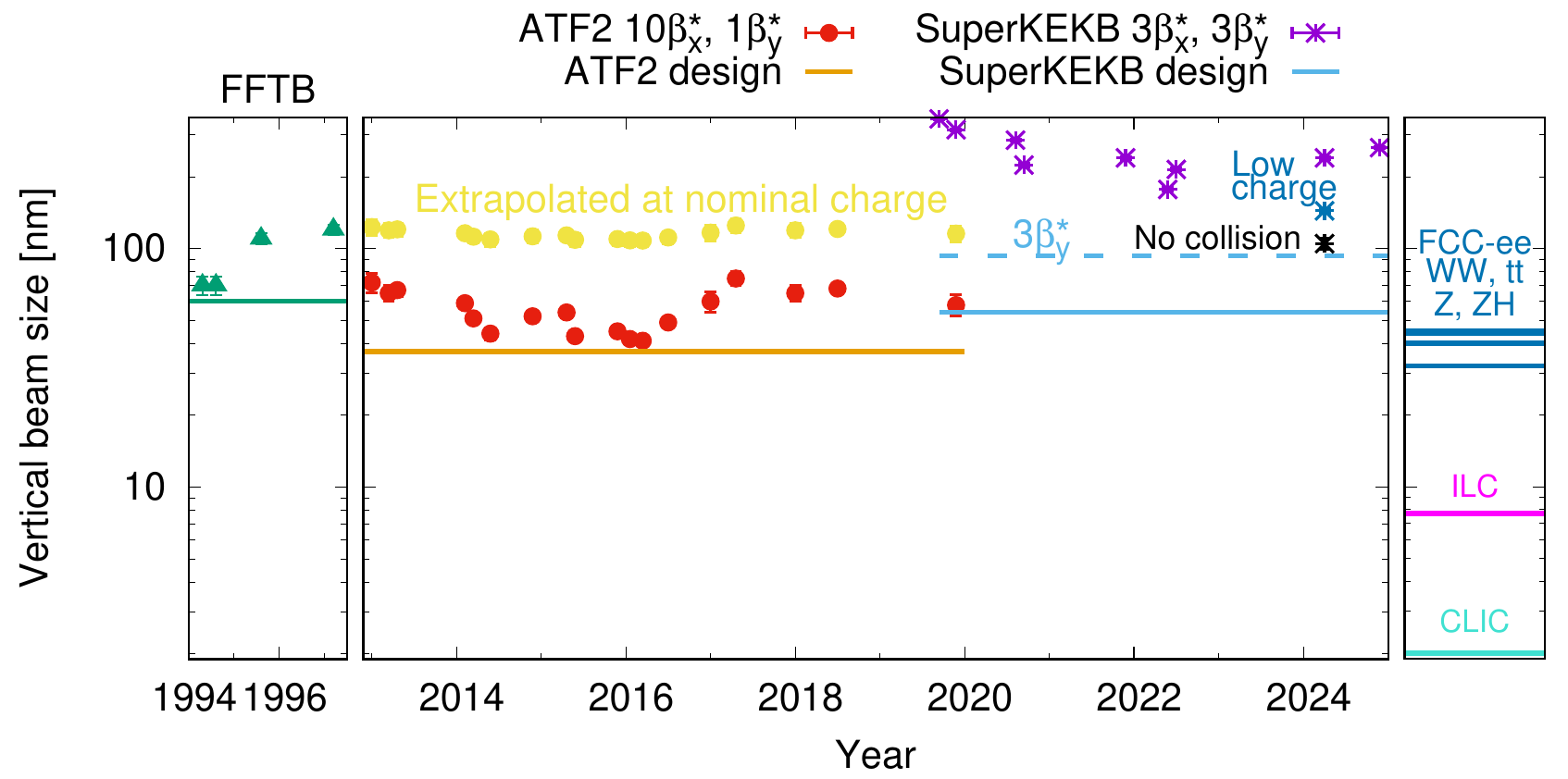}
\caption{\label{fig:sigmay_log} This is a version of Fig.~\ref{fig:sigmay} using a logarithmic scale for the vertical axis. }
\end{figure}

\begin{figure}[h]
\centering
\includegraphics[width=1\linewidth]{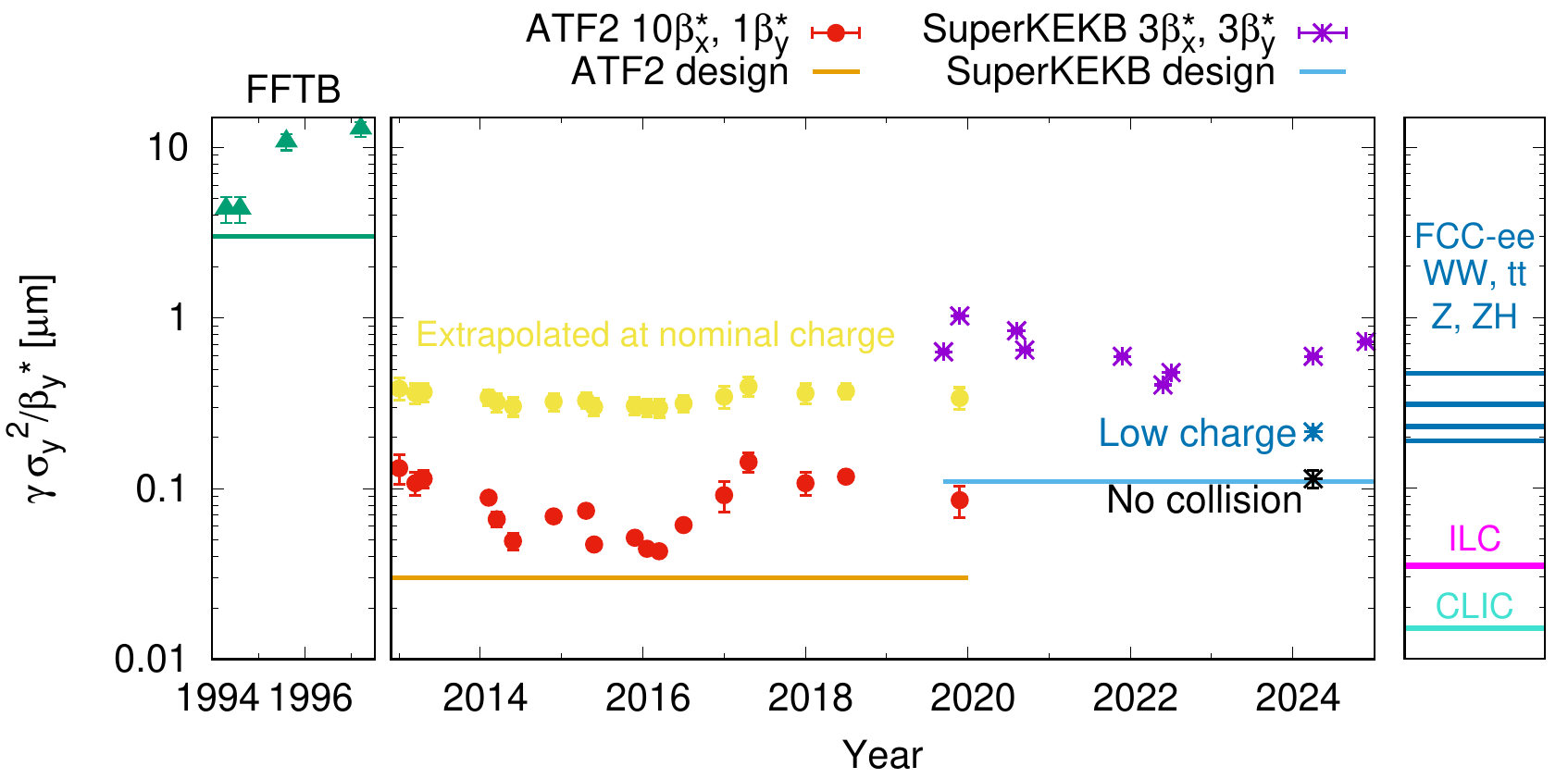}
\caption{\label{fig:emit} Effective normalized vertical emittance at the IP as measured from the beam size versus time. The projects and the color code are the same as in  Figs.~\ref{fig:sigmay} and \ref{fig:sigmay_log}. }
\end{figure}

Design parameters for these three projects are listed in Table~\ref{tab:design} and
vertical beam size measurements are shown in Fig.~\ref{fig:sigmay} versus time. Figure~\ref{fig:sigmay_log} shows another version of the plot with logarithmic vertical axis to improve visibility of the future projects with low beam size. 
An effective normalized vertical emittance at the IP as reconstructed from the measured beam size or the luminosity is also shown in Fig.~\ref{fig:emit}.
It is important to note that while the design vertical beam size for FFTB was 60~nm from~\cite{FFTBdesign} a lower value of 52~nm was quoted as the expected beam size in subsequent reports~\cite{FFTBprl}.
In addition, the beam energy of FFTB was estimated as 50~GeV in the design phase while it was operated at 45.6~GeV.

FFTB achieved 70$\pm6$~nm in the first years of operation~\cite{FFTBprl} using  lower 
bunch intensity (0.6$\times10^{10}$ e$^-$) and vertical emittance ($\gamma\epsilon_y=2\ \mu$m) than initially conceived in the design.
In shorter runs during later years this small beam size was not reproduced~\cite{LC97}.

ATF2 achieved a beam size of 65$\pm5$~nm in 2013~\cite{ATF2prl,okugi14} with a  reduced  
bunch population in the range of $0.13\times10^{10}$ to $0.21\times10^{10}$ electrons, 
and with a 10 times larger than design horizontal $\beta^*_x$. 
In 2016, a value of 41.1$\pm0.7$~nm was achieved~\cite{okugi} with an even lower bunch intensity of $0.07\times10^{10}$ e$^-$ and again a 10 times larger $\beta^*_x$. 
Figure~\ref{fig:sigmay} shows the time series of the smallest beam size over the ATF2 runs~\cite{okugi,patecki,yang20,yang21} in red.  
Using measurements of beam size ($\sigma_y$) versus bunch intensity ($N$) as reported in~\cite{yang20}:
\[
\sigma_y=\sqrt{\sigma_{y,0}^2 + w^2N^2}\,,\ \text{with } w=127\pm5 \text{ nm/nC}\,,
\]
it is possible to extrapolate the ATF2 measurements to nominal bunch intensity as shown on the figure in yellow.

SuperKEKB has been operating with  $\beta_{y}^\ast$ 
values of 0.8 and 1~mm, which correspond to 2.7--3.7 times the design values, and with a bunch charge about a factor 2 below design.
At SuperKEKB the vertical beam size at the collision point can be inferred from luminosity data, obtaining the geometrical average between the LER and HER beam sizes~\cite{demin}. 
Also, for each beam separately,  
the vertical emittance is measured with an X-ray monitor~\cite{xray}.  
The full time data series can be retrieved from the regular 
KEKB Accelerator Review Committee meetings~\cite{MAC}. 
Reducing bunch charge by about a factor of 6 from the design value the beam size decreased  by about 100~nm in 2024, see Fig.~\ref{fig:sigmay}. Furthermore, when measuring the emittance
with the X-ray monitor~\cite{xray}, the beam size at the collision point can also be estimated in the absence of collisions. 
Estimating IP beam size without collisions and also at low bunch charge shows a further reduction in the IP beam size to 105$\pm$6~nm, as indicated in black in Fig.~\ref{fig:sigmay}. The uncertainty for this data point is estimated from the observed systematic and random shifts when comparing X-ray monitor data and IP beam size data while in collisions.
It should also be noted that, presently, 
establishing collisions requires IP orbit changes followed by coupling and dispersion adjustments that increase equilibrium emittance.

\begin{table}
\hspace{-0.2cm}
\begin{tabular}{lcccccccccc}
    & FFTB  & ATF2 & \multicolumn{2}{c}{SuperKEKB} & \multicolumn{4}{c}{FCC-ee} & ILC & CLIC\\
    &       &      & LER & HER & Z & W & ZH & $t\overline{t}$ \\\hline
Particle     & e$^-$ &   e$^-$ & e$^+$ & e$^-$ & \multicolumn{4}{c}{e$^-$e$^+$} & e$^-$e$^+$ & e$^-$e$^+$\\  
Energy [GeV] & 50 
& 1.3 &  4 &  7 & 45.6 & 80 & 120 & 182.5 & 125 & 190\\
Charges [$10^{10}$] & 1-2 
 & 0.5 & 9 & 6.5 & 21.8 & 13.8 & 16.9 & 15.8 & 2 & 0.52\\
Emitt. $\gamma\epsilon_y$ [$\mu$m]  & 3 
& 0.03 & 0.068 & 0.177 & 0.19 & 0.31 & 0.23 & 0.47 & 0.035 & 0.015\\
IP drift (L*) [m]   & 0.4 & 1 & 0.765 & 1.22 & \multicolumn{4}{c}{2.2} & 3.5/4.3 & 6 \\
IP vert. $\beta^*_y$ [mm]  & 0.1 & 0.1 & 0.27 & 0.3 & 0.7 & 1 & 1  & 1.4  & 0.41 & 0.1 \\
IP vert. $\sigma_y$ [nm]  &60 
  & 37 & 48 & 62 & 40 & 45 & 32 & 44 & 7.7 & 2\\\hline
\end{tabular}
\caption{\label{tab:design} Design values of the different accelerator projects from~\cite{FFTBdesign,ATF2design,SuperKEKBdesign,FSR,ILC,CLIC}.}
\end{table}

\section{Summary}
ATF2, FFTB  and SuperKEKB have experimentally achieved vertical beam sizes of 41.1$\pm$0.7, 70$\pm$6 and  105$\pm$6~nm, respectively.
However, these projects show that achieving stable and repeatable operation at these low beam sizes remains a challenging endeavor.  It is observed that, so far, these projects did not achieve their design targets with nominal bunch intensity and that optics modifications  were adopted in ATF2 and SuperKEKB.
In particular, ATF2 increased its horizontal IP beta function by a factor 10 and SuperKEKB increased both IP beta functions by about a factor 3.

\section*{Acknowledgements}
We thank G.~Arduini, P.~Bambade, A.~Faus-Golfe, T.~Okugi, Y.~Papaphilippou, R. Scrivens and D.~Schulte for fruitful discussions.


\end{document}